\documentclass
[twocolumn,english,aps,superscriptaddress,floatfix,final,showpacs,prb,eqsecnum]{revtex4}%
\usepackage[T1]{fontenc}
\usepackage[latin1]{inputenc}
\usepackage{amsmath}
\usepackage{graphicx}
\usepackage{amssymb}
\usepackage{epsfig}
\usepackage{dcolumn}
\usepackage{bm}
\usepackage{amsfonts}
\usepackage{graphicx}
\usepackage{babel}%
\begin{document}
\title{RKKY interactions in the regime of strong localization}
\author{J. A. Sobota}
\affiliation{School of Applied and Engineering Physics, Cornell University, Ithaca, NY
14853, USA.}
\author{D. Tanaskovi\'{c}}
\affiliation{Institute of Physics, P.O. Box 57, 11080 Belgrade, Serbia.}
\affiliation{Department of Physics and National High Magnetic Field Laboratory, Florida
State University, Tallahassee, Florida 32306, USA.}
\author{V. Dobrosavljevi\'{c}}
\affiliation{Department of Physics and National High Magnetic Field Laboratory, Florida
State University, Tallahassee, Florida 32306, USA.}

\begin{abstract}
We study the influence of strong nonmagnetic disorder on the
Ruderman-Kittel-Kasuya-Yosida (RKKY) interactions between diluted magnetic
moments in metals. We find that the probability distribution for the RKKY
interactions assumes strongly non-Gaussian form featuring long tails. Since
such distributions cannot be characterized by its moments, we define a
\textit{typical} value of the interaction amplitude, which we find to be
exponentially suppressed in presence of Anderson localization. Our results
present a plausible and physically transparent picture describing how Anderson
localization effectively eliminates the long range nature of the RKKY interactions.

\end{abstract}

\maketitle



\section{Introduction}

It has long been appreciated that localized magnetic moments in metals
interact through indirect RKKY interactions mediated by the conduction
electrons.\cite{rkky} In a clean metal, the RKKY interaction has a long-range
oscillatory part, with an amplitude which decreases as a power law of the
distance between the impurities, $I(R)\sim\cos(2k_{F}R)/R^{d}$ ($d$ is the
dimensionality of the system and $k_{F}$ the Fermi wave vector). This behavior
is well understood to be a direct consequence of the existence of a sharp
Fermi surface characterizing itinerant electrons. \ \

In presence of impurities and disorder, this behavior may be substantially
modified. It is well known that sufficiently strong disorder can lead to
multiple-scattering processes which can trap the electrons through the
processes of Anderson localization. In this regime, one may expect that the
long-range character of RKKY interactions should be suppressed, reflecting the
reduced mobility of conduction electrons. The essential physical question is
how this process precisely takes place as the disorder strength is gradually
increased and the system approaches the localized regime.

The influence of weak nonmagnetic disorder on the RKKY interactions has been
studied in considerable
detail,\cite{chatel,degennes,zyuzin,bulaevskii,jagannathan} and is by now very
well understood. These studies have established that the main \ effect of weak
disorder is to randomly modify the phase of the RKKY oscillations due to
impurity-induced phase shifts of the electronic wave functions. As a result,
the RKKY interaction decreases exponentially when averaged over disorder,
$\langle I(R)\rangle\sim e^{-R/l}$, where $l$ is the mean free path. Early
work \cite{degennes} thus predicted that the range of RKKY\ interactions
becomes essentially cut off at $R\sim l$, and can be neglected at larger
distances. More careful consideration \cite{zyuzin,bulaevskii,jagannathan}
discovered that this naive argument is incorrect, essentially because most
relevant quantities do not depend on the average value of the interaction, but
instead on the typical value of its amplitude. Since this quantity is not
affected by the random phase shifts, it is essentially unaffected by weak
disorder, and decreases with the distance $R$ in the same power-law fashion as
in the clean system. Therefore, the presence of weak disorder is not expected
to lead to any significant changes in the physical properties which are
dominated by the RKKY interactions.

The influence of stronger nonmagnetic disorder on RKKY interactions has been,
however, so far poorly explored. In an important study, Lerner examined the
probability distribution of the RKKY interactions in the metallic phase in the
presence of strong disorder within generalized nonlinear $\sigma$ model and
$2+\varepsilon$ expansion.\cite{lerner} It was found that the quantum
interference corrections do not change the power-law decay of all the even
moments of the interaction distribution, which remains the same as in the pure
metal, but make the coefficients attached to these moments increase critically
with disorder. As the Anderson localization regime is approached, the higher
moments increase much faster than the variance, which therefore no longer
represents a good characterization of the typical interaction strength.
Although this work clearly points out to the importance of strong fluctuations
of the RKKY interactions, it does not explain how does the distribution
function evolve as one enters the regime of Anderson localization.

Within an Anderson insulator the electrons are bound to impurities, and thus
can hardly be expected to generate the long-range part of the RKKY
interaction. How can we have at the same time large moments of the
distribution of RKKY interactions, along with its fast decay with the
distance? At first glance these two arguments seem inconsistent and the
situation confusing and paradoxical. Clarifying these issues is an interesting
and important problem, since one expects the magnetic correlations to play a
crucial role in the physics of disorder-driven metal-insulator transitions in general.

The resolution of this puzzle is in fact quite simple, as we explain in this
paper. We find that in presence of strong disorder and localization, the
distribution function develops a strongly non-Gaussian form featuring long
tails. In such cases it is well known that all moments of the distribution can
assume very large values, while at the same time a typical width of the
distribution can remain very small. Instead of the arithmetic average (i.e.
standard deviation of the distribution) we find that the \textit{typical}
value of the interaction is better characterized by the geometric average of
the distribution, $I_{typ}(R)\equiv e^{\langle\frac{1}{2}\ln{[I(R)]^{2}%
}\rangle}$. This quantity is exponentially suppressed in the presence of
Anderson localization, explaining how the long-range part of the RKKY
interaction is suppressed.

To illustrate these ideas and obtain quantitative and reliable results, we
numerically study the distribution of the RKKY interactions within the
Anderson insulator phase. Although our numerical results are obtained within
one dimensional model, we argue that the same concept of the typical value of
the RKKY interaction can be used to physically explain the qualitative change
of the form of RKKY interactions from long ranged to short ranged during the
disorder driven metal-insulator transition in general dimensions. In the rest
of this paper we present a detailed description of the model, followed by the
numerical results and discussion.

\section{RKKY interactions in a disordered metal}

The interaction energy between two local moments ${\mathbf{S}}_{1}$ and
${\mathbf{S}}_{2}$ embedded in the metallic host at ${\mathbf{r}}_{1}$ and
${\mathbf{r}}_{2}$, respectively, is given by the Hamiltonian
\begin{equation}
\label{eq1}H_{int}=-J^{2}{\mathbf{S}}_{1} \cdot{\mathbf{S}}_{2} \,
\chi({\mathbf{r}}_{1},{\mathbf{r}}_{2}),
\end{equation}
where $\chi$ is the zero frequency nonlocal electronic susceptibility and $J$
is the exchange coupling constant. Therefore the calculation of the
interaction between diluted magnetic impurities reduces to the calculation of
the susceptibility which, expressed through the Matsubara Green's functions,
is given by
\begin{equation}
\label{eq2}\chi({\mathbf{r}}_{1},{\mathbf{r}}_{2})=\frac{2}{\beta}\sum
_{\omega_{n}}G_{\omega_{n}}({\mathbf{r}}_{1},{\mathbf{r}}_{2})G_{\omega_{n}%
}({\mathbf{r}}_{2},{\mathbf{r}}_{1}).
\end{equation}
Here $\omega_{n}$ is the fermionic Matsubara frequency, and $\beta$ is the
inverse temperature.

In the clean case, i.e.~in the absence of the nonmagnetic disorder, in three
dimensional (3d) system, and for $R$ much larger than the lattice spacing, the
susceptibility is equal to
\begin{equation}
\label{eq3}\chi_{o}(R)=-\frac{2mk_{F}\cos(2k_{F}R)}{(2\pi)^{3}R^{3}},
\end{equation}
where $R=|{\mathbf{r}}_{1}-{\mathbf{r}}_{2}|$ and $m$ is the effective mass.

In the presence of weak disorder the phase of $\chi(R)$ becomes random and
$\chi(R)$ averaged over the disorder configurations is exponentially
suppressed
\begin{equation}
\label{eq4}\langle\chi(R) \rangle=\chi_{o}(R)e^{-R/l},
\end{equation}
where $l$ is the mean free
path.\cite{chatel,degennes,zyuzin,bulaevskii,jagannathan} The second moment
(variance) of the probability distribution, however, remains long ranged and
has the same power law dependence as in the clean
system\cite{zyuzin,bulaevskii,jagannathan}
\begin{equation}
\label{eq5}\langle\chi^{2}(R) \rangle=3\left[  \frac{mk_{F}}{(2\pi)^{3}}
\right]  ^{2} \frac{1}{R^{6}}.
\end{equation}
Jagannathan \textit{et al.}\cite{jagannathan} have found that the square root
of the forth moment of the distribution, $\sqrt{\langle\chi^{4} \rangle}$, is
comparable in magnitude to the second moment. Therefore, the susceptibility
distribution is non-Gaussian, but its typical value is well characterized by
the square root of its second moment. The same remains true in $2d$ as
well.\cite{jagannathan}

This statement is, however, not valid in the regime of strong disorder. As
shown by Lerner,\cite{lerner} using the generalized nonlinear $\sigma$ model
and performing a  $2+\varepsilon$ expansion, further increase of the disorder
results in very rapid increase of all the even cumulants of the distribution.
More precisely, for the cumulant $\langle\langle\chi^{n}\rangle\rangle$ of the
order \nolinebreak$n$
\begin{equation}
\frac{\langle\langle\chi^{n}(R)\rangle\rangle}{R^{-nd}}\sim e^{2un^{2}%
},\label{eq6}%
\end{equation}
where the parameter $u\gtrsim1$ as the disorder is increased and the system
approaches the Anderson transition.\cite{lerner} In this case the typical
value of the distribution cannot be determined by the value of its moments.


\section{Numerical results}

In order to examine the form of the RKKY interactions in the regime of strong
disorder, we proceed to a numerical study. We consider a tight binding model
\begin{equation}
H=-t\sum_{\langle ij\rangle\sigma}(c_{i\sigma}^{\dagger}c_{j\sigma
}+\mbox{h.c.})+\sum_{i\sigma}\varepsilon_{i}c_{i\sigma}^{\dagger}c_{i\sigma
}\label{eq7}%
\end{equation}
with nearest neighbor hopping $t$ and on-site random potential $\varepsilon
_{i}$, which is distributed uniformly in the interval $[-W/2,W/2]$. We
calculate the interaction between the magnetic impurities embedded into this
system at distance $R$. From Eqs.~(\ref{eq1}) and (\ref{eq2}) we see that all
the information that we need is contained in the single particle Green's
functions $G_{\omega_{n}}({\mathbf{r}}_{1},{\mathbf{r}}_{2})$. In the matrix
notation the Green function
\begin{equation}
\hat{G}(\omega_{n})=(i\omega_{n}-\hat{H})^{-1},\label{eq8}%
\end{equation}
and the problem reduces to the numerical summation of the corresponding matrix
elements over Matsubara frequencies $\omega_{n}$ which will be done in the
zero temperature limit. In order to obtain good statistics with large number
of disorder realizations and to reduce the finite size effects, we concentrate
in the following to a one dimensional system.

The numerical results for the clean system, Fig.~{1}, reproduce the well known
oscillatory form of the electronic susceptibility with the power law decay
with the distance, $\chi(R)\sim\cos(2k_{F}R)/R$. \begin{figure}[t]
\begin{center}
\includegraphics[  width=2.3in,
keepaspectratio]{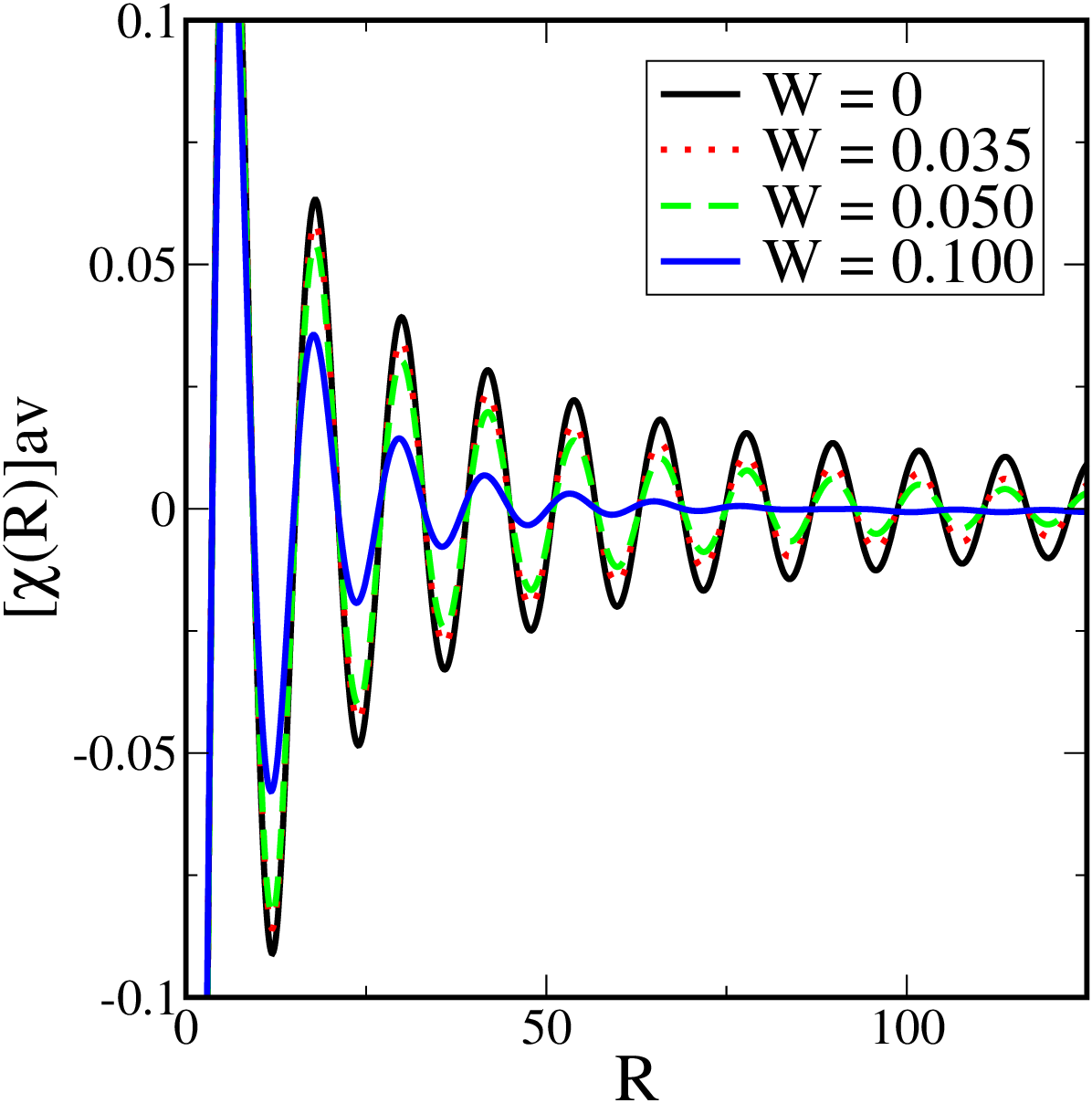}
\end{center}
\caption{The average susceptibility as a function of distance $R$ in the
presence of weak disorder of strength $W$. The chemical potential is $\mu =-0.96 $. }%
\end{figure}Here the power law exponent is equal to $1$ since we are working
within one dimensional (1d) model. (The disorder strength $W$ and
the chemical potential $\mu$ are measured in units of half the
bandwidth $2t$, and the distance $R$ in units of the lattice
spacing. Our system had $500$ lattice sites.) We then consider the
susceptibility $\langle\chi(R)\rangle$ in presence of weak
disorder averaged over hundreds of disorder configurations. The
average susceptibility weakens as the disorder is increased.
\begin{figure}[t]
\begin{center}
\includegraphics[  width=2.3in,
keepaspectratio]{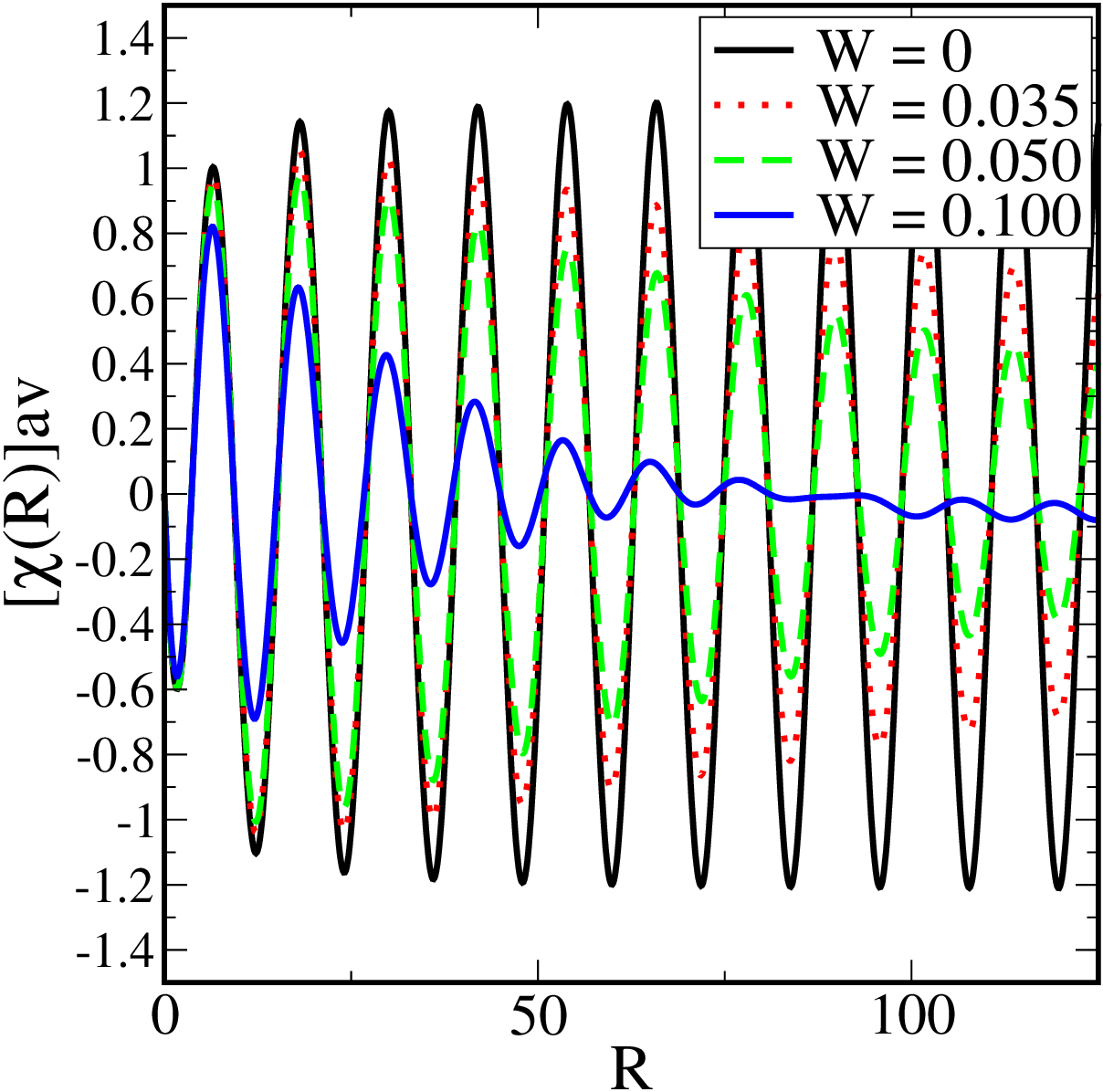}
\end{center}
\caption{Scaled average susceptibility $R\langle\chi(R)\rangle$. With the
$1/R$ dependence gone, we can see that disorder introduces a damping factor to
the average interaction strength.}%
\end{figure}

In Fig.~{2} we remove the $1/R$ dependence by multiplying the average value
$\langle\chi\rangle$ by $R$. For weak disorder $\langle\chi\rangle R$ follows
an exponential decay as predicted long time ago by de Gennes.\cite{degennes}
However, what we are really interested in are the probability distributions of
the electronic susceptibility in the presence of stronger disorder.
\begin{figure}[t]
\begin{center}
\includegraphics[  width=2.3in,
keepaspectratio]{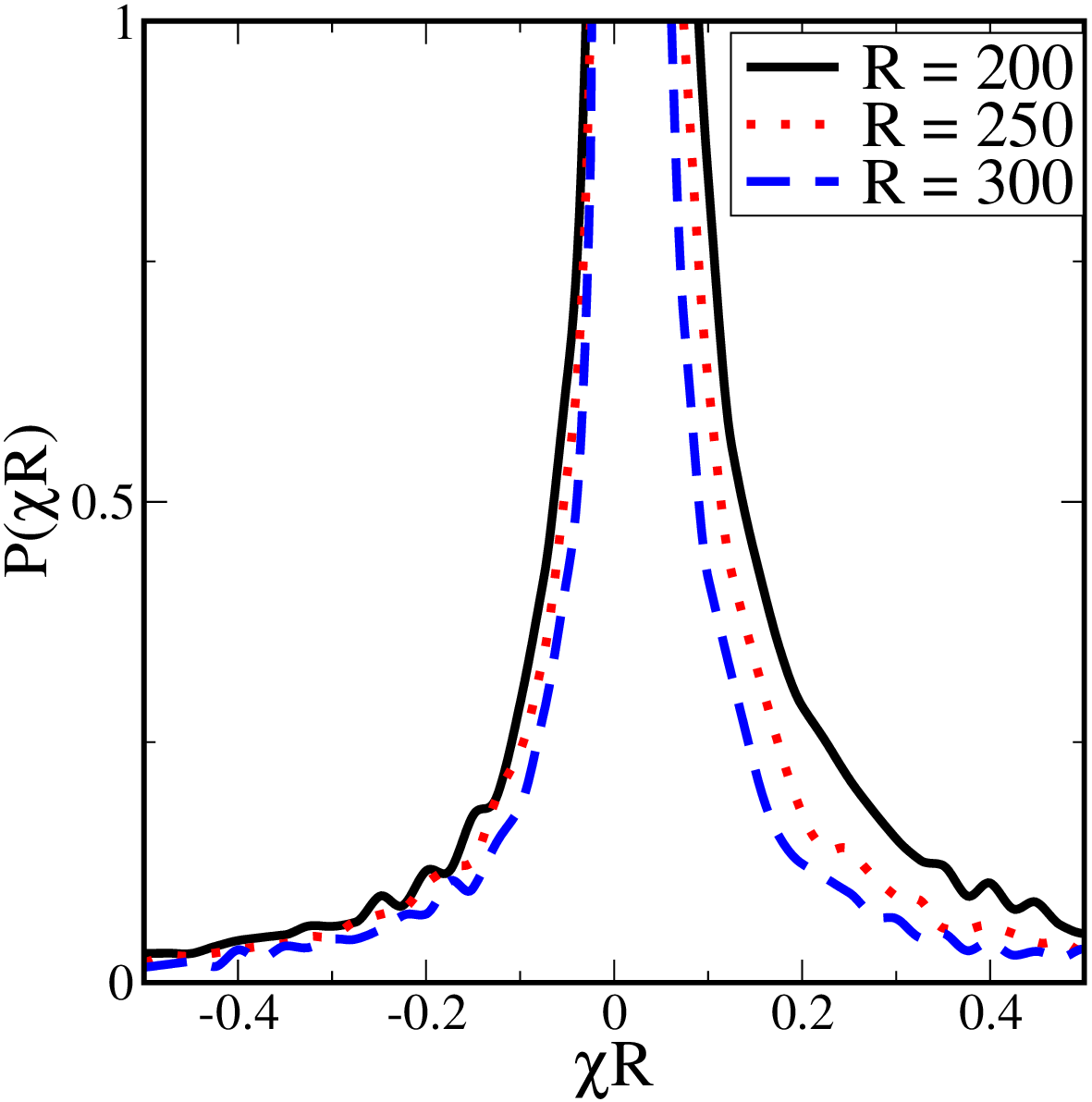}
\end{center}
\caption{Probability distribution $P(\chi R)$ for $W=0.35$ and
$\mu =-0.96$. The distribution width is clearly dependent on $R$.}
\label{fig3}%
\end{figure}

Fig.~{3 }shows the probability distribution of the scaled
susceptibility $P(\chi R)$ in the presence of strong disorder, and
for several values of $R$. As in the remaining part of the paper,
the results are obtained by averaging over hundreds of disorder
configurations on the lattice with $500-1000$ lattice sites. The
width of the distributions are dependent on $R$. More importantly,
cursory examination of the data indicates that the distributions
are distinctly non-Gaussian. This feature is important for the
following analysis, since the non-Gaussian shape of the
distribution prevents us from using the standard deviation as a
measure of the the width of the distributions. Instead, we define
a typical value of the width as the geometrical average
\begin{equation}
\chi_{typ}(R)\equiv e^{\langle\frac{1}{2}\ln\chi^{2}(R)\rangle}.\label{eq3.1}%
\end{equation}
\begin{figure}[t]
\begin{center}
\includegraphics[  width=2.2in,
keepaspectratio]{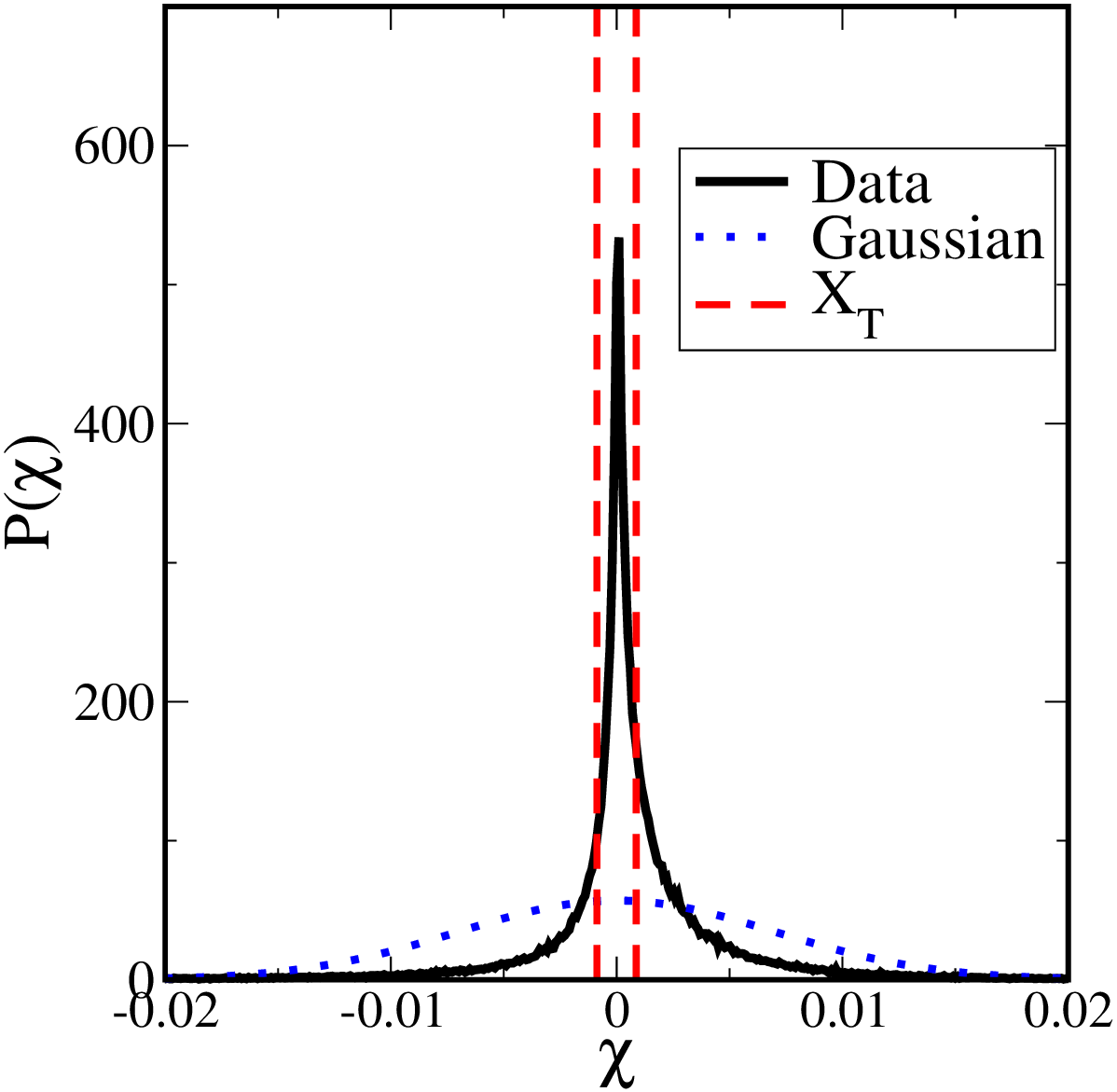}
\end{center}
\caption{Distribution $P(\chi)$ (full black line) for $R=100$ and
$W=0.2$. Gaussian (blue dotted line) is taken to have the same
standard deviation as $P(\chi)$. $\chi_{typ}$ (red dashed line) is
a better measure of the distribution width than
its standard deviation.}%
\end{figure}Fig.~4 shows a comparison of standard deviation and $\chi_{typ}$
for describing the width of the same distribution. We can see that the
standard deviation is influenced by the long tails, and is far too large to be
a useful description of the width. $\chi_{typ}$, however, gives a good
estimate of the width of the distribution.

Examining $\chi_{typ}$ for different strengths of disorder (Figs.~5 and 6), we
find that for each value of the disorder $W$
\begin{equation}
R\,\chi_{typ}(R)\sim e^{-R/\xi}\label{eq3.2}%
\end{equation}
for sufficiently large R, where $\xi$ defines the localization length.
\begin{figure}[t]
\begin{center}
\includegraphics[  width=2.3in,
keepaspectratio]{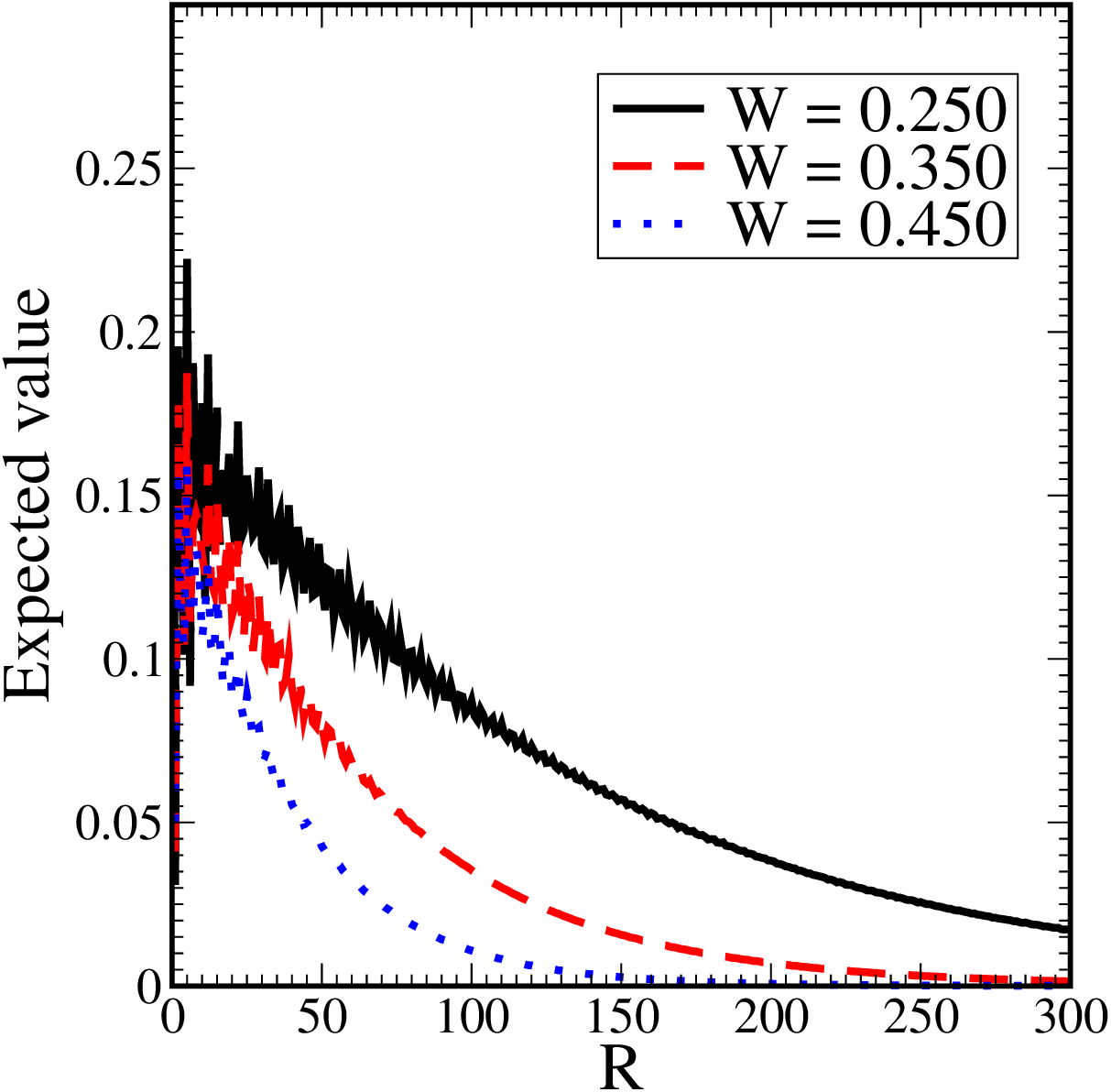}
\end{center}
\caption{$R\,\chi_{typ}$ on linear axes. The width of the distributions
decrease with increasing $R$.}%
\end{figure}\begin{figure}[t]
\begin{center}
\includegraphics[  width=2.3in,
keepaspectratio]{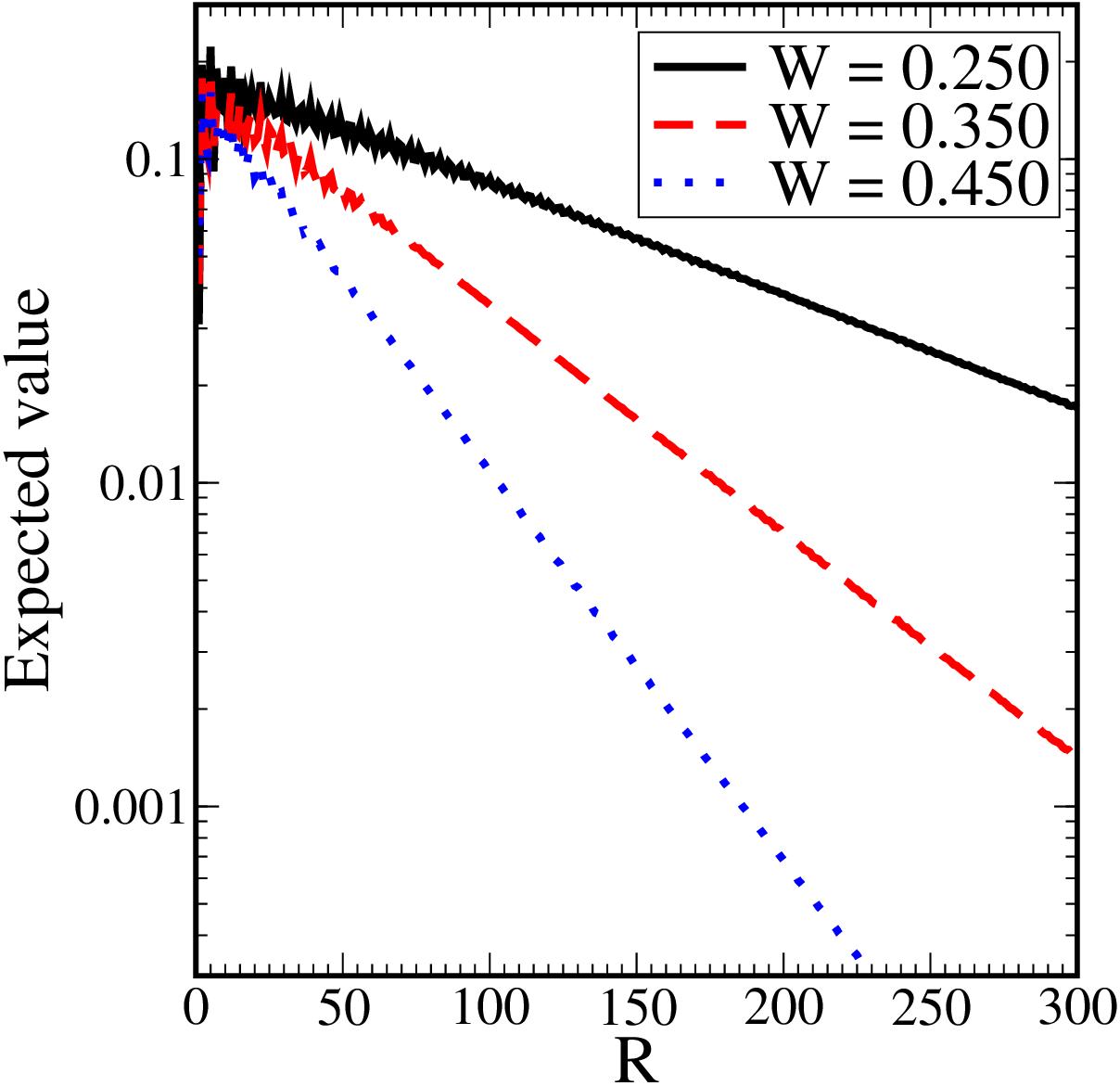}
\end{center}
\caption{$R\,\chi_{typ}$ on semilogarithmic axes. $R\,\chi_{typ}$ features an
exponential decay with $R$.}%
\end{figure}Therefore, if we define an adjusted susceptibility,
\begin{equation}
\chi_{A}(R)\equiv R\,e^{R/\xi}\chi(R)\label{eq3.3}%
\end{equation}
the distributions of $\chi_{A}(R)$ for large R will be entirely independent of
R within each disorder strength. \begin{figure}[h]
\begin{center}
\includegraphics[  width=2.3in,
keepaspectratio]{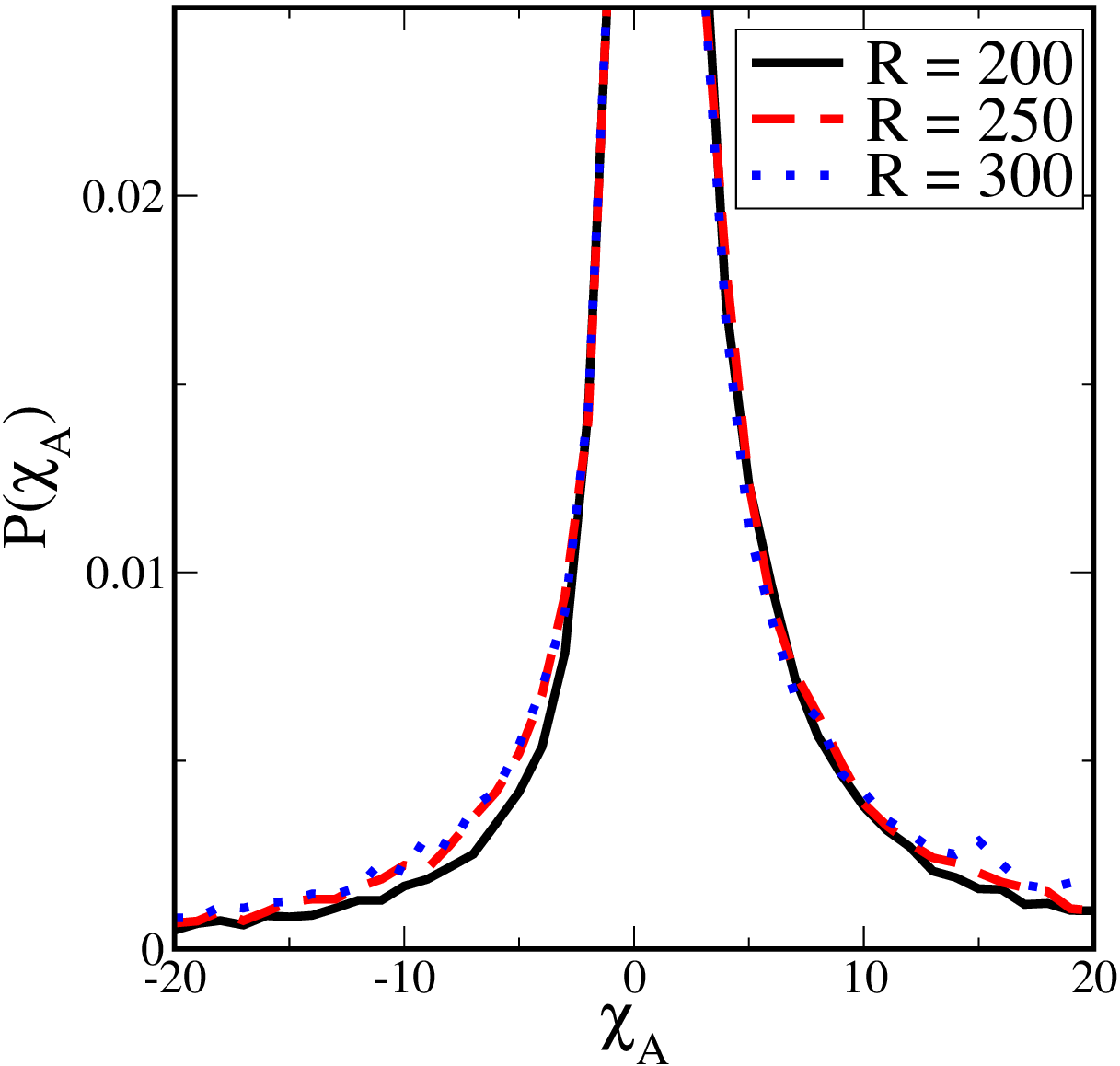}
\end{center}
\caption{Scaled $P(\chi_{A})$ for W = 0.350. After scaling, the distributions
collapse.}%
\end{figure}
\begin{figure}[h]
\begin{center}
\includegraphics[  width=2.3in,
keepaspectratio]{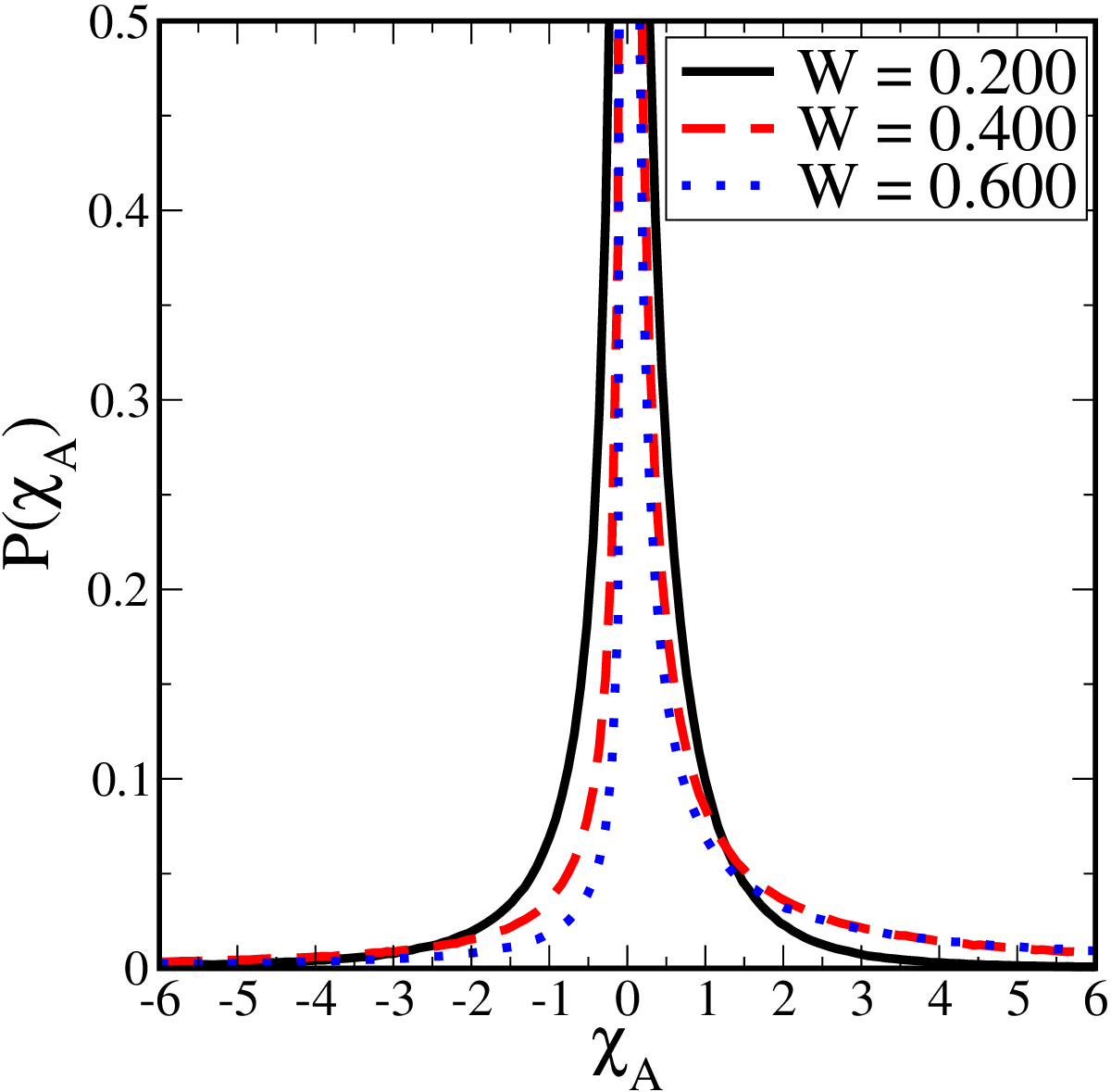}
\end{center}
\caption{$P(\chi_{A})$ for several values of $W$. The
distributions are
strongly peaked for higher disorder.}%
\end{figure}
Comparing Figs.~3 and 7 shows how the use of this adjusted
susceptibility causes these distributions to collapse to a single
scaling function. We then combine the data for several distances
$R$, and thereby obtain more precise distributions for each value
of disorder $W$. For each disorder strength there now is a single
characteristic distribution (independent on $R$) as shown in
Fig.~8. Interestingly, the probability distributions are quite
asymmetric in the presence of stronger disorder, ferromagnetic
interactions being much more probable than the antiferromagnetic
ones. In the strongly localize regime we expect in fact the
interactions to be ferromagnetic for R smaller or of the order of
the localization length. The obtained distributions show this
behavior even for R much larger that the localization length.

We then plot the tails of the distributions and find that for very
strong disorder they become very long,  as shown in Fig.~9. The
tails of the distribution appear to converge to a universal power
law form in the limit of strong disorder. The form of the tails is
qualitatively the same for positive and negative side of the
distribution.
\begin{figure}[h]
\begin{center}
\includegraphics[  width=2.3in,
keepaspectratio]{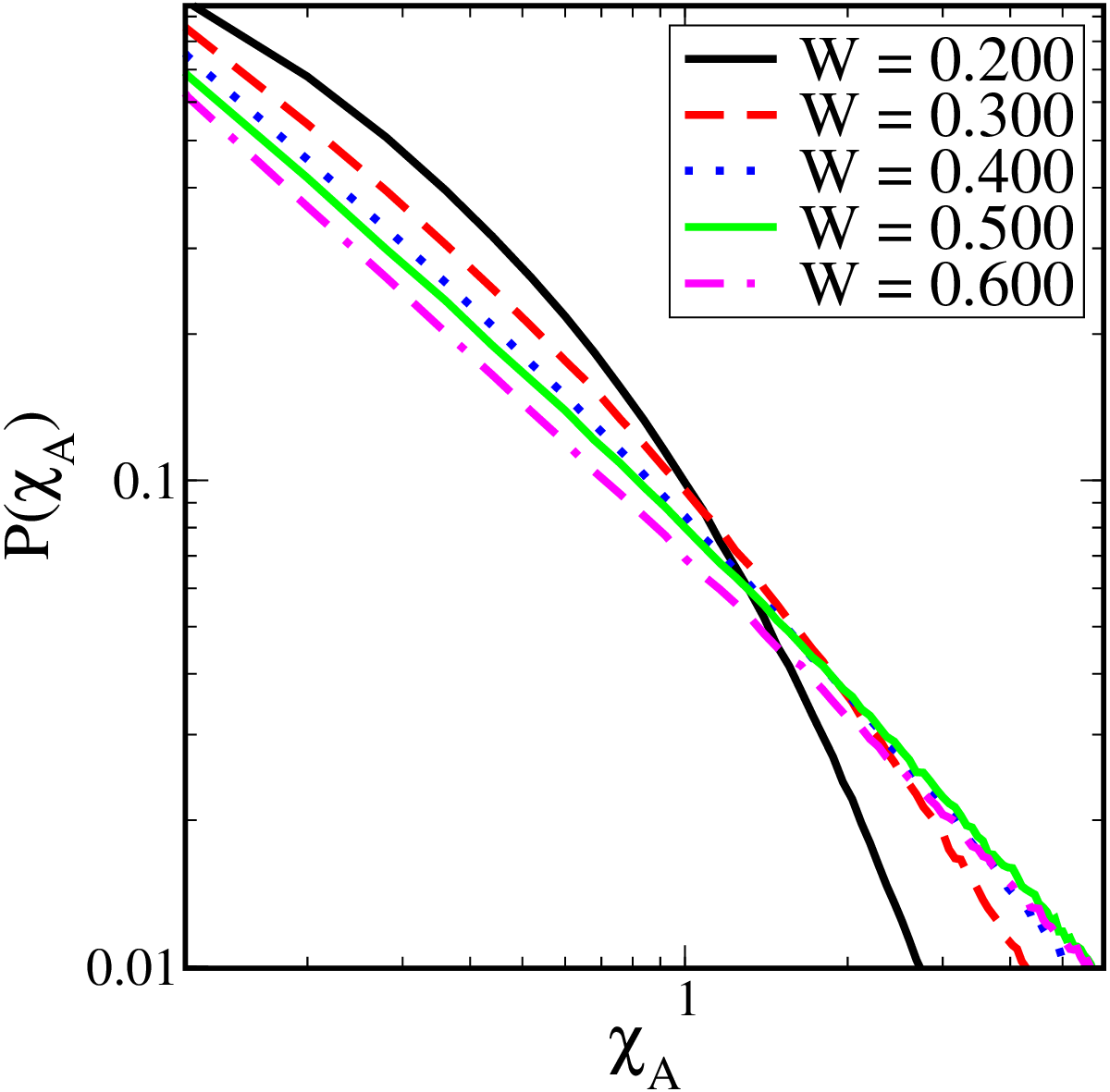}
\end{center}
\caption{Tails of distributions on log-log axes. For higher disorder, the
tails become increasingly long.}%
\end{figure}
\begin{figure}[h]
\begin{center}
\vspace*{.5cm} \includegraphics[  width=2.3in,
keepaspectratio]{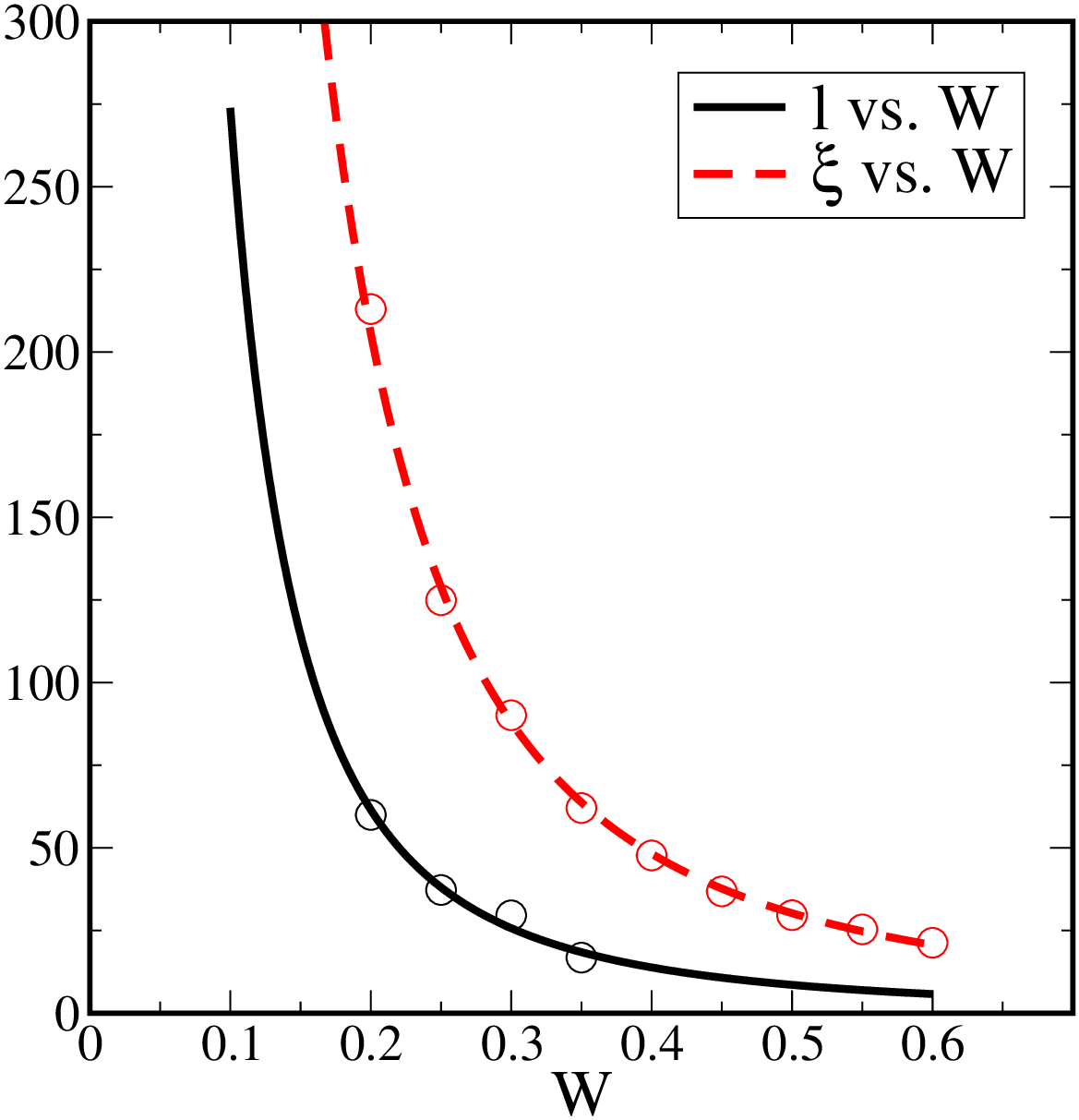} \label{fig11}
\end{center}
\caption{The mean free path $l$ and the localization length $\xi$
as a function of the disorder strength. Both $l$ and $\xi$ are
found to be proportional to $W^{-2}$.}%
\end{figure}
The existence of such long tails indicates that the width of the
distributions for higher disorder cannot be accurately
characterized by their moments. The moments of such a distribution
are extremely large, while the typical value is, in fact, very
small. Therefore, the long range part of the RKKY interactions is
strongly suppressed in the strongly localization regime.

We have also compared our result for the mean free path $l$, obtained from
Eq.~(\ref{eq4}), and the localization length $\xi$, from Eq.~(\ref{eq3.2}).
Plotting $l$ and $\xi$ as a function of $W$, see Fig.~10, we find that $l$ and
$\xi$ are both proportional to $W^{-2}$, and find that  $\xi/l\approx3.685$,
which is in a good agreement with the analytical result $\xi/l\approx
4$.\cite{economou} This analysis further confirms  the consistency of our
interpretation of $\xi,$ as determined from the decay of the typical RKKY
interaction amplitude, with the localization length of the electronic system.

\section{Conclusion}

In this paper we have examined how the distribution function for RKKY
interactions becomes modified due to Anderson localization effects. We
demonstrated that the essential effect of localization is to exponentially
suppress the typical amplitude of RKKY interactions on distances longer then
the localization length, in agreement with intuitive expectations. The
distribution, nevertheless, remains \textquotedblleft broad\textquotedblright%
\ in the sense that it develops long tails which dominate the statistics. Our
numerical results thus confirm the analytical predictions of Lerner that all
even moments of the distribution diverge within an Anderson insulator.

Our results portray an interesting physical picture with
potentially far-reaching consequences. In the metallic regime the
RKKY interactions remain long-ranged even in presence of weak
disorder, and thus a given magnetic moment effectively interacts
with many others. In this regime one may expect a well developed
collective behavior of the spin system, leading to magnetic
ordering at low temperature. When the Anderson--localized regime
is approached, the situation is quite different. The RKKY
interaction between a typical pair of distant spins is now
significantly suppressed or even negligibly small. Very
occasionally, a pair of distant spins will interact strongly, due
to rare disorder configurations producing long tails in the
distribution function. If the resulting RKKY interaction is
antiferromagnetic, then such a pair can be expected to lock in a
tightly bound singlet -- thus forming an essentially inert
molecule that practically detaches from the rest of the spin
system. Such process are precisely what one expects within the
random singlet phases
\cite{bhattlee81,dasgupta,paalanenbhatt,sarachik} which feature
quantum Griffiths phase anomalies.\cite{eduardovladreview}.
Alternatively, a ferromagnetic interaction will lead to the
formation of a bound triplet state, essentially a spin $S=1$
magnetic moment. It is interesting to note a degree of asymmetry
of the distribution of the RKKY interactions, that we have found
at strong disorder. This finding seems to indicate that
ferromagnetic correlations may effectively compete with the
tendency for singlet formation, possibly leading to nano-scale
ferromagnetism\cite{nielsen} coexisting with a random singlet
phase.

We thus anticipate that Anderson localization processes generically
destabilize spin glass ordering in disordered metals, which is instead
replaced by an appropriate quantum Griffiths phase. Precisely how these
processes take place in the vicinity of realistic metal-insulator transitions
remains a fascinating open direction for future study.

\section{Acknowledgements}

We thank E. Abrahams, R. Bhatt, S. Chakravarty, E. Miranda, and B. Spivak for
useful discussions. This work was supported by the National High Magnetic
Field Laboratory, NSF grants DMR-0234215 (V.D.~and D.T.) and DMR-0542026
(V.D.), Center for Integrating Research and Learning stuff (J.A.S.), and
Serbian Ministry of Science, Project No.~141014 (D.T.).

\end{document}